\documentclass[11pt]{article}

\usepackage[preprint]{acl}

\usepackage{times}
\usepackage{latexsym}

\usepackage[T1]{fontenc}

\usepackage[utf8]{inputenc}

\usepackage{microtype}

\usepackage{inconsolata}

\usepackage{graphicx}

\usepackage{xspace}
\usepackage{graphbox}
\usepackage{multirow}
\usepackage{tablefootnote}
\usepackage{wrapfig}
\usepackage{amsmath}
\usepackage{booktabs}
\usepackage{amssymb}
\usepackage{tabularx}
\usepackage{enumitem}
\usepackage{makecell}
\usepackage{listings}
\usepackage{hyperref, url}
\usepackage{graphicx,booktabs,array,adjustbox}
\usepackage[capitalize,noabbrev]{cleveref}

\usepackage{tikz}
\usetikzlibrary{positioning,fit,backgrounds,arrows.meta,calc}
\colorlet{cadd}{green!55!black}\colorlet{caddbg}{green!14}%
\colorlet{crem}{red!70!black}\colorlet{crembg}{red!12}%
\colorlet{cctx}{blue!60!black}\colorlet{cctxbg}{blue!9}%
\colorlet{csite}{orange!80!black}\colorlet{csitebg}{orange!16}%
\colorlet{cana}{violet!70!black}\colorlet{canabg}{violet!8}%
\providecommand{\stepnum}{}\renewcommand{\stepnum}[1]{\tikz[baseline=(c.base)]\node[circle,fill=black!75,text=white,inner sep=0.5pt,minimum size=2.4mm,font=\sffamily\bfseries\tiny](c){#1};}%
\pgfdeclarelayer{frames}%
\pgfsetlayers{background,frames,main}%

\newcolumntype{R}[2]{>{\adjustbox{angle=#1,lap=\width-(#2)}\bgroup}l<{\egroup}}

\newcommand{\sysrb}{\textsc{SysRB}}
\newcommand{\parabf}[1]{\noindent\textbf{#1}~\ignorespaces}

\title{Do Coding Agents Understand Why Systems Crash---and Why They Don't?}

\author{
  \textbf{Chenxi Huang\textsuperscript{1}},
  \textbf{Haonan Wang\textsuperscript{1}},
  \textbf{Jinghe Zhang\textsuperscript{1}},
  \textbf{Aleksandr Nogikh\textsuperscript{2}},
  \textbf{Franjo Ivan\v{c}i\'{c}\textsuperscript{2}},
\\
  \textbf{Petros Maniatis\textsuperscript{3}},
  \textbf{Eugene Wu\textsuperscript{1}},
  \textbf{Kostis Kaffes\textsuperscript{1}},
  \textbf{Junfeng Yang\textsuperscript{1}},
  \textbf{Baishakhi Ray\textsuperscript{1}}
\\
\\
  \textsuperscript{1}Columbia University,
  \textsuperscript{2}Google Inc.,
  \textsuperscript{3}Google DeepMind
\\
  \small{
    \textbf{Correspondence:} \href{mailto:chenxi@cs.columbia.edu}{chenxi@cs.columbia.edu}
  }
}

\begin{document}
\maketitle


\begin{abstract}
Coding agents are increasingly capable of fixing real-world bugs, yet they still struggle with large systems codebases such as the Linux kernel.
To understand why, we test how agents reason about execution in these codebases through crash reasoning: predicting from source code whether a given input crashes the system.
A correct prediction alone, however, does not establish this ability, since it may come from merely spotting suspicious code.
In this paper, we propose source interventions: we pair source trees that differ by a single commit, so that the same input crashes only one of them, and test whether agents predict crashes more often on the crashing tree.
Experiments with four coding agents on \sysrb, our dataset of 934 such pairs from Linux and OSS-Fuzz projects, yield three main findings.
First, agents are biased toward predicting no crash: they rarely raise false alarms but miss many real crashes, and even the best agent misses over a quarter of them.
Second, locating the code that determines the crash outcome is not enough: agents often misjudge its effect, especially when the crash occurs far from that code.
Third, testing only on crashing code is misleading: agents that merely notice suspicious code predict crashes on both source trees, which would look like correct reasoning on the crashing tree alone.
Together, these findings indicate that current coding agents cannot yet reliably reason about execution in complex systems codebases, and that paired evaluation is necessary to measure this ability.
\end{abstract}
\section{Introduction}
\label{sec:intro}

Coding agents are increasingly used for diverse software engineering (SE)
tasks, including code generation, issue resolution, and test generation
~\citep{yang2024sweagent,rastenis2026automated,ryan2024code}.
However, reliable SE requires more than generating code:
agents must understand what happens when a program runs,
including how different inputs and code modifications affect its behavior.
This capability, known as \emph{execution reasoning}, is particularly
important for debugging and program repair, where identifying
the conditions that trigger a failure is essential to fixing it correctly.

Yet, how well coding agents perform execution reasoning remains unclear,
particularly in complex, reliability-critical system codebases.
For example, a manual study of 79 patches generated by kAgent,
an agent designed to repair Linux kernel crashes, found that 52\%
altered or removed intended functionality~\citep{mathai2026kagent}.
This raises a broader concern for critical system codebases:
fixing a failure without understanding its cause may introduce
unintended behavior.
More importantly, a patch that stops a crash does not show whether an agent understands the execution that caused it.
We therefore need to evaluate execution reasoning independently
of code generation or repair success.

Existing code-reasoning benchmarks attempt to measure this capability.
Early benchmarks focus on predicting outputs, program semantics, or
static-analysis properties of individual functions and code
fragments~\citep{gu2024cruxeval,roy2026codesense,xie2026core}.
Recent benchmarks extend these evaluations to repositories, examining execution traces, call chains, and intermediate program states~\citep{wang2026reporeasoner,
li-etal-2026-laboratory,taherkhani2026llmsreasonruntimebehavior}. 
However, these benchmarks either focus on small pieces of code
or require tracking program execution step by step,
neither of which scales well to large, complex system codebases. 
For example, Linux kernel execution may involve multiple subsystems,
hardware interactions, and concurrent threads, making comprehensive
execution tracing expensive and difficult~\citep{308764}.
We therefore need a way to evaluate program understanding
without tracking every step of execution.

However, not all execution outcomes are equally difficult to evaluate.
One particularly useful outcome is whether a program crashes on a given input,
which can be verified by running the program.
Motivated by this observation, we study \emph{crash reasoning}:
given a program's source code, build configuration, and an input,
an agent must predict whether the program will crash and, if so,
where and why.
This allows us to evaluate execution reasoning in complex system codebases
without having to reconstruct every step of program execution.

Crash reasoning also arises naturally in practical debugging scenarios.
For example, in security testing, automated fuzzing systems such as
syzbot~\citep{syzbot} and OSS-Fuzz~\citep{serebryany2017oss} generate large
numbers of inputs to identify those that cause programs to crash.
Developers must then determine why these inputs trigger crashes
and modify the code to prevent them
~\citep{317997,bohme2017bug,bai2026beyond}.

However, a correct crash prediction alone does not necessarily imply that
an agent understands why the crash occurs.
For example, an agent may identify a suspicious operation and predict
a crash without determining whether the given input actually triggers it.
Such a prediction could be correct for a crashing program but remain
unchanged after the crash is resolved.
Evaluating only the crashing source tree would therefore misrepresent the agent's reasoning ability.

\begin{figure*}[t]
\centering
\vspace{-8pt}
\resizebox{0.95\linewidth}{!}{\includegraphics[]{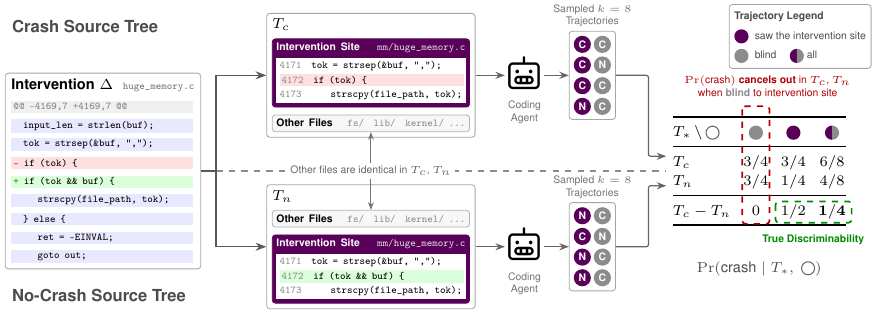}}
\caption{\textbf{Differencing over a pair with opposite crash outcomes isolates true
discriminability.} $T_c$ and $T_n$ differ only by $\Delta$ at the intervention site. When sampling coding agent to predict ``crash or not'' multiple times, trajectories blind to the site predict a crash equally often in
both trees and cancel in $T_c - T_n$.}
\label{fig:cr-overview}
\vspace{-16pt}
\end{figure*}

To address this limitation, we introduce \emph{source interventions}
(\Cref{fig:cr-overview}).
Instead of evaluating an agent on a single program, we compare its predictions on two source trees of the same program: one that crashes for a given input and another that does not.
The trees differ by one code change; we run both on the same input to verify their outcomes. The agent examines each tree independently, without knowing which crashes.
We evaluate its behavior along two dimensions:


\noindent (i) \textbf{Crash-outcome discriminability.}
We measure how often an agent predicts a crash on each source tree.
An agent that makes the same prediction on both source trees cannot
distinguish their execution outcomes, even if its prediction
is correct on one of them.
The difference in crash-prediction rates therefore measures
the agent's sensitivity to code changes that affect program
behavior.


\noindent (ii) \textbf{Reasoning behavior.}
We examine the agent's investigation steps to check whether it finds and analyzes the differing code (\emph{intervention site}) and the code where the crash occurs (\emph{crash site}).
This helps us understand whether agents fail to locate relevant
code or struggle to predict its effect on execution even after
encountering it.


\parabf{Findings.}
We instantiate this evaluation in \textsc{SysRB} (\textbf{Sys}tem \textbf{R}easoning \textbf{B}enchmark), a dataset of 934 execution-validated source-tree pairs from the Linux kernel and OSS-Fuzz projects, and evaluate four coding agents.
First, agents frequently miss real crashes despite low false-positive rates: they incorrectly predict no crash for $27.5$--$76.8\%$ of crashing trees, while predicting crashes for only $2.8$--$10.9\%$ of non-crashing trees (\Cref{sec:eval:overall}).
Second, finding relevant code does not guarantee correct outcome discrimination.
Although agents observe the intervention site in $55.6$--$89.0\%$ of trajectories, their discriminability conditioned on observing it ranges from only $21.5\%$ to $75.4\%$.
For Claude Opus 5, discriminability further drops from about $65\%$ to $38\%$ when the crash occurs in a different subsystem from the intervention site (\Cref{sec:eval:decomposition}).
Third, single-source-tree evaluations can reward misleading predictions.
When Opus analyzes the crash site without observing the intervention site, its crash-prediction rate increases by about 37 percentage points on $T_c$, but also by 34 points on $T_n$.
Similar patterns occur across all four agents (\Cref{sec:eval:trajectories}).
Together, these findings reveal limitations in agents' ability to distinguish execution outcomes and demonstrate the value of paired evaluation for studying system-level code reasoning.

\parabf{Our contributions.}
We make three contributions:
\begin{itemize}[noitemsep, topsep=0pt,leftmargin=*]
    \item \textbf{Source-intervention-based evaluation.}
    We introduce a paired evaluation methodology that measures whether coding agents distinguish crash outcomes across execution-validated source interventions.
    Unlike single-source-tree evaluation, this design reduces the influence of prediction tendencies shared across crashing and non-crashing programs.

    \item \textbf{An execution-validated benchmark and diagnostic framework.}
    We construct~\sysrb, comprising 934 source-tree pairs from the Linux kernel and OSS-Fuzz projects.
    We complement policy-level discriminability with trajectory-level analysis of whether agents observe and analyze the intervention and crash sites.

    \item \textbf{Empirical insights into agent limitations.}
    Evaluating four coding agents, we find that agents frequently miss crashes, that observing relevant code does not guarantee correct outcome discrimination, and that cross-subsystem failures are associated with substantially lower discriminability.
    We demonstrate how single-source-tree evaluations can reward predictions driven by suspicious code patterns shared across both source trees.
\end{itemize}

\section{Related Work}
\label{sec:related}

\parabf{Crash resolution.}
ARVO~\citep{mei2026arvo}, kGym~\citep{mathai2024kgym}, and Live-kBench~\citep{huang2026outrunning} provide reproducible fuzzer-found crashes in open-source libraries and the Linux kernel, on which agents have been built to resolve OSS-Fuzz~\citep{zhang2026fixing} and Linux kernel crashes~\citep{singh2025code, mathai2026kagent}.
Suppressing a crash, however, does not by itself demonstrate an understanding of its mechanism~\citep{mathai2026kagent, huang2026outrunning}.
We instead evaluate \emph{crash reasoning} directly, independently of patch generation.

\parabf{Code reasoning.}
Function-level benchmarks test input–output prediction~\citep{gu2024cruxeval,xu2025cruxeval} and finer-grained semantics such as runtime values, branch outcomes, and dependencies~\citep{roy2026codesense, xie2026core, 10.1109/ICSE55347.2025.00038};
repository-level benchmarks examine execution traces, intermediate states, and call chains from test runs~\citep{wang2026reporeasoner,li-etal-2026-laboratory}.
Both rely on isolated functions or recorded intermediate states, which do not transfer readily to system codebases.
Through \emph{source interventions}, we advance the code reasoning to system-level software by evaluating crash reasoning and measure \emph{crash-outcome discriminability}.

\section{Analyzing Agentic Crash Reasoning}
\label{sec:method}

\subsection{Crash Reasoning with Source Intervention}
\label{sec:causal-model}
\label{sec:agentic-crash-reasoning}

\parabf{Program crash.}
A program built from a source tree $T$ under a build configuration executes a sequence of operations on an input, each of which reads or modifies program state.
If execution reaches a crash condition, the crash then occurs and the crash handler generates a crash report.
We call the operations and state transitions that lead to the crash the \emph{crash mechanism}, and an input known to trigger a crash a \emph{reproducer}; reproducers are found manually or by fuzzing.

\parabf{Execution oracle.}
Let $x$ denote a reproducer together with its build configuration, and let the \emph{execution oracle} $\operatorname{Exec}(T,x)$ return the outcome of building $T$ under that configuration and running the reproducer on the resulting program:
\[
\operatorname{Exec}(T,x)=(y,C),\quad y\in\{\textsf{crash},\textsf{no-crash}\}
\]
where $C$ is the crash report if $y=\textsf{crash}$, and $C=\varnothing$ when a valid execution reports no crash.
The report gives the crash type and a stack trace to the \emph{crash site}, where the crash occurs (Appendix~\ref{appendix:crash-report}).
We repeat each execution to confirm a stable outcome (\Cref{sec:method:pair-instance-curation}).

\parabf{Crash-reasoning task.}
A coding agent predicts $\mathrm{Exec}(T,x)$ from source code.
Given a source tree $T$ and reproducer $x$, the agent returns a verdict $v\in\{\textsf{crash},\textsf{no-crash}\}$ and, when $v=\textsf{crash}$, a predicted crash report $\widehat C$ with the crash type and a stack trace to the crash.
The agent may navigate the repository, search and inspect files, and run Bash commands, but cannot access the Internet, compile the source, or execute the reproducer.
It receives no crash report or indication of which code is relevant, so it must infer the outcome from source code.
We model the agent as a policy $\pi$ that investigates and reasons over its observations:
\[
(\tau,v,\widehat C)\sim\pi(\cdot\mid T,x),
\]
where $\tau$ records tool calls, returned observations, and explicit reasoning.

However, crash-verdict accuracy on crashing instances alone cannot distinguish crash reasoning from a tendency to predict \textsf{crash}.
If every instance crashes, an agent that always answers \textsf{crash} and names a suspicious-looking crash site would appear correct, and agents do produce such verdicts as we show in \Cref{sec:eval:trajectories}.
We therefore use pairs of $(T, x)$ that differ only slightly and \emph{look alike} to an agent, yet have opposite crash outcomes.
An agent that distinguishes their execution outcomes should predict \textsf{crash} more often on the crashing tree.
We construct these pairs by intervening on source tree $T$ while holding the reproducer $x$ fixed.

\parabf{Source intervention.}
To obtain the opposite crash outcome, we can change either the reproducer $x$ (a reproducer intervention) or the source tree $T$ (a source intervention); we intervene on the source for two reasons.
First, real source interventions are abundant and tied to each crash: for crashes found by fuzzing, the commit that fixes the crash or the commit that bisection identifies as its cause is routinely recorded \citep{syzbot, serebryany2017oss}.
Second, a real commit ties the opposite outcome to the crash's root cause and keeps the practitioner's task intact: given a reproducer from a fuzzer, find the code that determines whether it crashes.
Formally, an intervention applies a single commit $\Delta$ to a base tree $T^-$, producing $T^+ = T^- \oplus \Delta$, while keeping $x$ and the execution environment fixed.
We retain pairs whose outcomes differ and denote the crashing and non-crashing trees by $T_c$ and $T_n$:
\[ \resizebox{\columnwidth}{!}{$\mathrm{Exec}(T_c,x)=(\textsf{crash},C),\quad \mathrm{Exec}(T_n,x)=(\textsf{no-crash},\emptyset).$} \]
$\Delta$ either removes the crash (a fix commit, $T_c = T^-$) or introduces it (a cause-bisected commit, $T_c = T^+$).
The \emph{intervention site} is the lines changed by $\Delta$ and their surrounding context.

\parabf{Nuisance differences.}
We suppress residual cues such as file modification times (\Cref{sec:method:pair-instance-curation}) to ensure $\Delta$ as the only perceptible difference between $T_c$ and $T_n$.
As a descriptive check, we also report discriminability for trajectories that never observe the intervention site (\Cref{sec:eval:env-noise}).

\parabf{Evaluation approach.}
Since $T_c$ and $T_n$ differ only by $\Delta$, we evaluate agents at two levels.
At the policy level, we measure crash-outcome \emph{discriminability} (\Cref{sec:policy-evaluation}): an agent that reasons about the crash mechanism should predict a crash more often on $T_c$ than on $T_n$.
At the trajectory level, we examine agents' reasoning behavior (\Cref{sec:trajectory-evaluation}).
We inspect the turns in each trajectory where the agent reads and analyzes the intervention site and the crash site, to determine whether the agent fails to find the relevant code or fails to reason about the execution behavior of the code found.
\Cref{sec:method:pair-instance-curation} describes how we construct the pairs in \sysrb.

\begin{figure*}[t]
\centering
\vspace{-0.8\baselineskip}
\resizebox{0.95\textwidth}{!}{\includegraphics[]{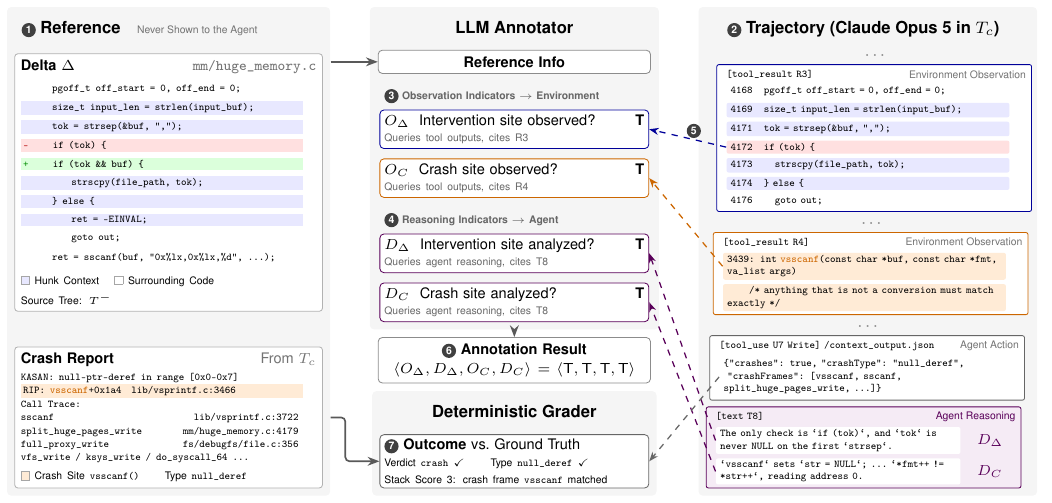}}
\caption{Trajectory-level evaluation on one instance (Linux fix \texttt{893eaef1}, source tree $T_c$, Claude Opus 5)}
\label{fig:pipeline}
\vspace{-12pt}
\end{figure*}

\subsection{Policy-Level Evaluation}
\label{sec:policy-evaluation}

To evaluate the policy's discriminability to source interventions, we sample verdicts on both source trees of each instance and estimate the probability of predicting a crash.
We run each agent $K$ times independently on each tree and estimate its crash-prediction probability as the fraction of runs that predict \textsf{crash}:
\[
\widehat q_i^a
=
\frac{1}{K}\sum_{r=1}^{K}
\mathbf{1}\!\left[v_{i,a}^{(r)}=\textsf{crash}\right].
\]
where $v_{i,a}^{(r)}$ is the verdict of run $r$ on tree $T_a$ of instance $i$.
Averaging $\widehat q_i^c-\widehat q_i^n$, an estimate of Youden's $J$, over the $N$ instances gives the agent's \emph{discriminability}
\begin{align*}
J(\pi)
&=
\frac{1}{N}\sum_{i=1}^{N}(\widehat  q_i^c - \widehat q_i^n)
\end{align*}
i.e., the true-positive rate minus the false-positive rate, with \textsf{crash} as the positive class.
$J$ ranges from $-1$ to $1$: $J=1$ means perfect discriminability, with every verdict \textsf{crash} on $T_c$ and \textsf{no-crash} on $T_n$; guessing \textsf{crash} with equal likelihood on both trees contributes zero in expectation, even when it is right on one of them; and negative values mean the agent responds opposite to the execution effect.

\subsection{Trajectory-Level Evaluation}
\label{sec:trajectory-evaluation}

Although chain-of-thought quality is hard to measure directly, we inspect trajectory $\tau$ to check whether the agent reads and reasons about the intervention and crash sites.
This helps identify failure patterns in agentic crash reasoning.
We also score the similarity between predicted and reference crash stack frames with a deterministic grader.

\parabf{Site profile.}
An LLM annotator labels each trajectory $\tau$ with four indicators (\Cref{fig:pipeline}):
\[
Z(\tau)=(O_\Delta,D_\Delta,O_C,D_C).
\]
The \emph{observation} indicators record whether the site is reached by the agent: $O_\Delta$ holds when a tool output shows the intervention site, and $O_C$ indicates whether code from the crash-site function body is shown to the agent.
The \emph{reasoning} indicators record whether the agent analyzed that site: $D_\Delta$ for reasoning about execution at the intervention site and $D_C$ for the crash-site semantics.
The annotator receives $\tau$, the prediction $(v,\widehat C)$, $\Delta$, $C$, and whether the tree is $T^-$ or $T^+$ (\Cref{fig:pipeline}\,\stepnum{1}), but not whether it is $T_c$ or $T_n$, to reduce bias from knowing whether the verdict is correct.
Every \textsf{T} label must cite verbatim text from $\tau$, rendered as numbered blocks that separate tool outputs from the agent's text (\stepnum{2}--\stepnum{4}); a string-matching checker marks labels with unverifiable citations as ungraded (\stepnum{5}).
Therefore a site profile is returned (\stepnum{6}).

\parabf{Stack trace similarity.}
Since a \textsf{crash} verdict on $T_c$ counts toward $J$ even if it names a different crash, we also compare the predicted crash type with the reference and score the predicted stack trace with a deterministic grader (\Cref{tab:stack-rubrics}) (\stepnum{7}).

\begin{table}[t]
\centering
\small
\setlength{\tabcolsep}{4pt}
\begin{tabular}{@{}cl@{}}
\toprule
Score & Predicted stack trace (highest-score criterion met) \\
\midrule
3 & Crash frame matches the reference \\
2 & An adjacent caller--callee edge matches \\
1 & Crash frame lies in the reference subsystem \\
0 & Otherwise \\
\bottomrule
\end{tabular}
\caption{Stack-trace score against the reference report.}
\label{tab:stack-rubrics}
\vspace{-16pt}
\end{table}

\subsection{Dataset Curation}
\label{sec:curation}
\label{sec:method:pair-instance-curation}

We instantiate this paired design in \textsc{SysRB}~(\textbf{Sys}tem \textbf{R}easoning \textbf{B}enchmark), whose split \texttt{sysrb.v1} contains 934 execution-validated pairs: 545 Linux kernel crashes from syzbot~\citep{syzbot} and 389 open-source system library crashes found by OSS-Fuzz~\citep{serebryany2017oss}, drawn from the ARVO dataset~\citep{mei2026arvo}.
Each instance ships the reference report from our reproduction on $T_c$, $T_c$ and $T_n$ as Docker images built from the same base image, with git history removed and filesystem metadata randomized so that $\Delta$ is the only intended difference, and without compilers or build tools; it also includes the reproducer but excludes the reference crash report.

\begin{table*}[t]
\centering
\small
\setlength{\tabcolsep}{3.5pt}
\begin{tabular*}{\textwidth}{@{\extracolsep{\fill}}l rr rrr rrr rrr@{}}
\toprule
& \multicolumn{2}{c}{Mean Per Run} & \multicolumn{3}{c}{Full} & \multicolumn{3}{c}{Linux} & \multicolumn{3}{c}{OSS-Fuzz} \\
\cmidrule(lr){2-3} \cmidrule(lr){4-6} \cmidrule(lr){7-9} \cmidrule(lr){10-12}
Model & Turns & Cost (\$) & $J$ & TPR & FPR & $J$ & TPR & FPR & $J$ & TPR & FPR \\
\midrule
DeepSeek V4 Flash & 50.2 & -- & $+12.3 \pm 1.9$ & 23.2 & 10.9 & $+6.5 \pm 1.8$ & 20.3 & 13.8 & $+20.5 \pm 3.5$ & 27.3 & 6.8 \\
GLM 5.3 & 85.8 & -- & $+37.6 \pm 2.6$ & 41.7 & 4.0 & $+26.6 \pm 2.9$ & 29.9 & 3.3 & $+53.1 \pm 4.0$ & 58.3 & 5.1 \\
Gemini 3.8 Flash & 188.7 & 2.94 & $+46.2 \pm 3.0$ & 49.0 & \textbf{2.8} & $+37.2 \pm 3.6$ & 40.2 & \textbf{3.0} & $+60.0 \pm 4.4$ & 62.5 & \textbf{2.5} \\
Claude Opus 5 & 63.1 & 6.39 & $\textbf{+67.7} \pm 2.7$ & \textbf{72.5} & 4.8 & $\textbf{+62.3} \pm 3.6$ & \textbf{67.8} & 5.4 & $\textbf{+75.3} \pm 4.0$ & \textbf{79.2} & 3.9 \\
\bottomrule
\end{tabular*}
\vspace{-0.8\baselineskip}
\caption{Pooled discriminability of prevalent coding agents on \sysrb~(\%); $\pm$ is a 95\% bootstrap CI over instances.}
\label{tab:j-by-model}
\vspace{-16pt}
\end{table*}

\parabf{Linux split (\texttt{sysrb.v1.linux}).}
Using the LiveKBench curator~\citep{huang2026outrunning}, we collect syzbot crashes with a syzkaller reproducer, the crash's kernel configuration, and either a fix commit ($T_c=T^-$) or a cause-bisected commit ($T_c=T^+$).
We build both trees and run the reproducer 25 times on each in kGymSuite~\citep{mathai2026kagent}, retaining pairs in which $T_c$ yields a report whose title matches the syzbot crash and $T_n$ yields none.
After removing unstable pairs, 545 of the 779 validated pairs remain.

\parabf{OSS-Fuzz split (\texttt{sysrb.v1.oss}).}
ARVO provides reproducible OSS-Fuzz crashes with Docker images for vulnerable and fixed revisions.
ARVO records only fix commits, so $T_c=T^-$ for all instances.
To ensure $\Delta$ is a single commit, we select the 582 crashes whose vulnerable revision is the fix commit's parent. We retain the 389 for which re-execution confirms a stable crash on $T_c$ and none on $T_n$.

\section{Evaluation}
\label{sec:eval}

In this section, we study how agents perform on crash-reasoning tasks.
We report discriminability and stack-trace prediction performance in \Cref{sec:eval:overall}, break discriminability down by crash type, and characterize the overall agent behaviors.
We then decompose discriminability $J$ over trajectories with different site profiles.
In \Cref{sec:eval:decomposition}, we study the relationship between discriminability and the intervention site, and further analyze performance on crashes where the intervention site is far from the crash site.
In \Cref{sec:eval:trajectories}, we identify the bias revealed by source interventions and how much the bias affects evaluation metrics on $T_c$ alone, on $T_n$ alone, and in the paired setting.


\subsection{Experiment Setup}
\label{sec:eval:setup}

\parabf{Coding agents.}
We evaluate four coding agents on \sysrb:
Claude Opus 5, GLM 5.3~\citep{zeng2026glm}, and DeepSeek V4 Flash (0731)~\citep{xu2026deepseek} on Claude Code, and Gemini 3.8 Flash on Google Antigravity CLI.
Each agent runs for up to two hours with an unlimited number of turns, and we sample $K=8$ trajectories per instance and source tree.
Following \Cref{sec:policy-evaluation}, we use Youden's $J$ to measure intervention discrimination at the policy level, and site profiles to characterize evidence grounding at the trajectory level (\Cref{sec:trajectory-evaluation}).
Appendix~\ref{appendix:model-params} details the serving and sampling configuration.

\parabf{LLM annotator.}
We use Gemini 3.8 Flash to annotate trajectories with $(O_\Delta,D_\Delta,O_C,D_C)$ per the criteria in \Cref{fig:pipeline}, and score predicted stack traces deterministically according to \Cref{tab:stack-rubrics}.
We manually check $160$ annotations sampled uniformly at random from all experiments and find $93.1\%$ agreement with our labels.

\subsection{Overall Crash-Reasoning Performance}
\label{sec:eval:overall}

\parabf{Discriminability varies widely across agents and crash types.}
\Cref{tab:j-by-model} shows that $J$ ranges from $12.3\%$ for DeepSeek V4 Flash to $67.7\%$ for Claude Opus 5, and that all four agents discriminate better on OSS-Fuzz than on Linux.
The gap between agents on Linux is wider than OSS-Fuzz.
In Linux split, Opus reaches $62.3\%$, $25.1$ pp. above the next-best agent, Gemini, compared with a $15.3$ pp.\ lead on OSS-Fuzz.
DeepSeek falls to $6.5\%$ on Linux, predicting a crash on $13.8\%$ of non-crashing trees against $20.3\%$ of crashing ones.
Besides, as shown in \Cref{tab:j-by-crash-type}, $J$ also varies in different crash type.
For Opus on Linux, hung-task and deadlock reports, which arise from interactions across threads rather than from a single faulting operation, reach only $39.5\%$ and $40.5\%$, against $68.5\%$ for KASAN and $71.7\%$ for UBSAN reports.
On OSS-Fuzz, $J$ varies even among memory-safety crash types, from $37.5\%$ for double free to $81.4\%$ for heap buffer overflows.

\begin{table}[b]
\centering
\scriptsize
\setlength{\tabcolsep}{2pt}
\vspace{-8pt}
\begin{tabularx}{\columnwidth}{@{}>{\raggedright\arraybackslash}X r@{\hspace{1.2em}}>{\raggedright\arraybackslash}X r@{}}
\toprule
\multicolumn{2}{@{}c}{\textbf{Linux}} & \multicolumn{2}{c@{}}{\textbf{OSS-Fuzz}} \\
\cmidrule(r{1.2em}){1-2}\cmidrule{3-4}
Crash type & $\overline{J}$ & Crash type & $\overline{J}$ \\
\midrule
hung task \emph{(19)}                & +39.5 & double free \emph{(10)}           & +37.5 \\
deadlock \emph{(46)}                 & +40.5 & stack buffer overflow \emph{(27)} & +61.0 \\
divide error \emph{(7)}              & +43.5 & global buffer overflow \emph{(9)} & +64.9 \\
BUG \emph{(89)}                      & +56.4 & uninitialized value \emph{(58)}   & +66.6 \\
WARNING \emph{(167)}                 & +63.9 & SEGV \emph{(50)}                  & +74.6 \\
NULL deref \emph{(87)}               & +66.3 & heap use-after-free \emph{(45)}   & +74.9 \\
KASAN \emph{(88)}                    & +68.5 & heap buffer overflow \emph{(173)} & +81.4 \\
UBSAN \emph{(23)}                    & +71.7 & index out of bounds \emph{(5)}    & +100.0 \\
general protection fault \emph{(19)} & +80.3 & other \emph{(8)}                  & +71.9 \\
\bottomrule
\end{tabularx}
\caption{Mean per-crash $J$ (\%) for Claude Opus 5, by crash type with at least five instances}
\label{tab:j-by-crash-type}
\vspace{-8pt}
\end{table}

\begin{figure}[t]
    \centering
    \includegraphics[width=\linewidth]{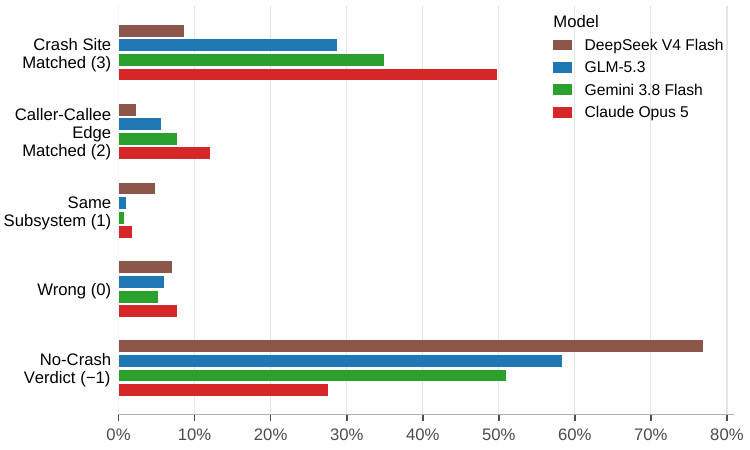}
    \caption{Frequency (\%) of stack trace score on $T_c$}
    \label{fig:score-dist}
    \vspace{-8pt}
\end{figure}


\parabf{Agents tend toward predicting no crash.}
Although \Cref{tab:j-by-model} shows positive discriminability for all four agents, their errors are strongly asymmetric: false-negative rates ($1 - \text{TPR}$) range from $27.5\%$ to $76.8\%$, whereas FPR range from $2.8\%$ to $10.9\%$.
Agents thus rarely raise false alarms but miss many real crashes; even Opus misses more than a quarter of them.
Opus's lead over Gemini comes from recognizing more crashes (TPR $72.5\%$ vs.\ $49.0\%$) despite higher false alarms (FPR $4.8\%$ vs.\ $2.8\%$).

\parabf{When agents predict a crash, they usually locate the crash correctly.}
As shown in \Cref{fig:score-dist}, conditional on predicting \textsf{crash} on $T_c$, GLM, Gemini, and Opus identify the correct crash site in most of the cases, while DeepSeek does fewer than half of its crash predictions.
This contrasts with their frequent missed crashes: although the agents often fail to recognize that a crash occurs, their positive predictions usually identify where it occurs.

\subsection{Discriminability and the Intervention Site}
\label{sec:eval:decomposition}

\begin{figure}[t]
  \centering
  \includegraphics[width=\linewidth]{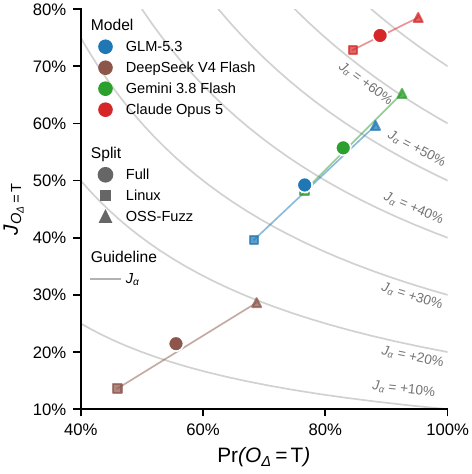}
  \caption{$J_{O_\Delta{=}\mathrm{T}}$ versus $\Pr(O_\Delta=\mathsf{T})$}
  \label{fig:decomp-odelta}
  \vspace{-12pt}
\end{figure}

\noindent We now condition discriminability on the site profile $Z(\tau)$.
Starting with the observation indicator $O_\Delta$, we separate how often agents observe the intervention site from how well they discriminate \emph{conditional on} observing it.
Let $p=\Pr(O_\Delta=\mathsf{T})$, averaged over $T_c$ and $T_n$, be the intervention-site observation rate, and let $J_{O_\Delta{=}\mathsf{T}}$ and $\epsilon$ denote discriminability among trajectories that do and do not observe the intervention site.
Because the observation rate differs by at most $3.3$ pp.\ between $T_c$ and $T_n$ for every agent and split, $J$ decomposes as
$$
J \approx \underbrace{p\,J_{O_\Delta{=}\mathsf{T}}}_{J_\alpha} + \underbrace{(1-p)\,\epsilon}_{J_\beta}
$$
with error at most $0.6$ pp.\ in all our experiments (Appendix~\ref{app:j-decomposition}).
$J_\alpha$ is the contribution of trajectories that observe the intervention site, and $J_\beta$ that of trajectories that do not.

\parabf{Finding intervention site does not ensure discrimination.}
\Cref{fig:decomp-odelta,tab:decomp-odelta} show differences among agents.
Opus has the highest intervention-site observation rate ($p = 89.0\%$) and conditional discriminability ($J_{O_\Delta{=}\mathsf{T}} = 75.4\%$).
GLM and Gemini find intervention site in $76.6\%$ and $82.9\%$ of trajectories, yet their conditional discriminability is only $49.2\%$ and $55.7\%$.
DeepSeek is lower on both measures, at $55.6\%$ intervention-site observation rate and $21.5\%$ conditional discriminability.
Even after observing this code, Opus predicts a crash on $79.7\%$ of $T_c$ trajectories (\Cref{tab:decomp-odelta-full}), leaving a gap from perfection.
The gap to perfect discrimination therefore comes both from missing intervention site and conditional discriminability.

\begin{table}[t]
  \small
  \setlength{\tabcolsep}{3.5pt}
  
\resizebox{\columnwidth}{!}{
\centering
\small
\setlength{\tabcolsep}{2pt}
\begin{tabular}{l ccccc}
\toprule
Model & $p$ & $J_{O_\Delta{=}\mathsf{T}}$ & $\epsilon$ & $J_\alpha$ & $J_\beta$ \\
\midrule
DeepSeek V4 Flash         & 55.6 & $+21.5$ & $+0.6$ & $+11.9$ & $+0.3$ \\
GLM 5.3                   & 76.6 & $+49.2$ & $-0.6$ & $+37.7$ & $-0.1$ \\
Gemini 3.8 Flash          & 82.9 & $+55.7$ & $-0.3$ & $+46.2$ & $-0.0$ \\
Claude Opus 5             & \textbf{89.0} & $\mathbf{+75.4}$ & $+1.9$ & $\mathbf{+67.1}$ & $+0.2$ \\
\bottomrule
\end{tabular}
}

  \caption{Discriminability decomposition ($\%$)}
  \label{tab:decomp-odelta}
  \vspace{-12pt}
\end{table}

\parabf{Without the intervention site, agents do not discriminate.}
\label{sec:eval:env-noise}
Although we remove known cues to source-tree identity (\Cref{sec:agentic-crash-reasoning}), an agent could still exploit an unanticipated side channel and discriminate without reasoning about the intervention site.
$\epsilon$ captures any such effect together with sampling noise: on the full dataset it ranges from $-0.6$ to $+1.9$ pp., and the corresponding $J_\beta$ from $-0.1$ to $+0.3$ pp (\Cref{tab:decomp-odelta}).
Virtually all measured discriminability therefore comes from trajectories that observe the intervention site.

\begin{figure}[t]
    \centering
    \includegraphics[width=\linewidth]{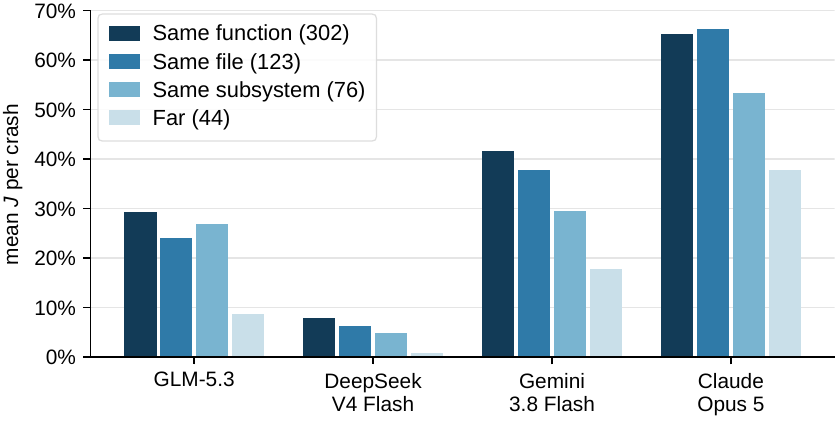}
    \caption{Mean per-crash $J$ on Linux by distance between the intervention and crash sites.}
    \label{fig:j-delta-dist}
    \vspace{-8pt}
\end{figure}

\begin{figure}[t]
    \centering
    \includegraphics[width=\linewidth]{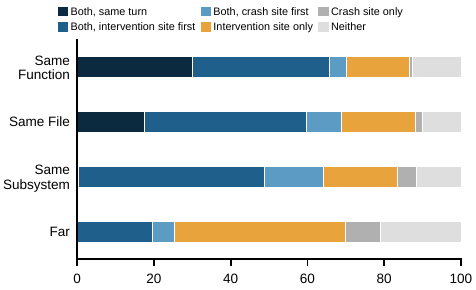}
    \caption{Order of observing intervention and crash sites in $T_c$ trajectories (Linux, Claude Opus 5).}
    \label{fig:order-breakdown}
    \vspace{-8pt}
\end{figure}

\parabf{Discrimination drops when the crash occurs far from the intervention site.}
We group Linux instances by the distance between the reference crash site and the intervention site: in the same function, the same file, the same subsystem, or far (a different subsystem).
All four agents reach their lowest $J$ in the far group (\Cref{fig:j-delta-dist}); for Opus, $J$ falls from about $65\%$ in the same-function and same-file groups to $38\%$ in the far group.
\Cref{fig:order-breakdown} breaks down the trajectories in each distance group based on the ordering between observations of intervention site and crash site on $T_c$.
Opus observes both sites in $70.1\%$, $69.0\%$, and $64.1\%$ of trajectories in the same-function, same-file, and same-subsystem groups, but only $25.2\%$ in the far group.
Instead, $44.7\%$ of far-group trajectories observe only the intervention site, compared with $16.5$--$19.4\%$ in the other groups, and $9.2\%$ observe only the crash site.
When the two sites first appear at different turns, the intervention site comes first three to eight times as often as the crash site ($35.7\%$ vs.\ $4.5\%$ of trajectories in the same-function group, and $19.5\%$ vs.\ $5.7\%$ in the far group).
Therefore, when the crash manifests in another subsystem, Opus usually finds the intervention site but often does not reach the crash site, a pattern that coincides with the lower $J$ of the far group.

\subsection{Source Interventions Reveal Bias}
\label{sec:eval:trajectories}

Without observing the intervention site, a trajectory lacks direct evidence of the source change; \Cref{sec:eval:decomposition} shows that such trajectories have near-zero discriminability.
They still produce verdicts, however, and whichever verdict they give is correct on one of the two trees.
We use the site profile to examine these verdicts and compare their influence to single-tree metrics and to $J$.


\parabf{Agents favor \textsf{no-crash} without observing the intervention site.}
On $T_n$, between $11.3\%$ and $44.0\%$ of correct \textsf{no-crash} verdicts come from trajectories without observing intervention site (\Cref{tab:nocrash-decomp}).
Among trajectories that do not observe the intervention site, $86.7\%$--$97.4\%$ predict \textsf{no-crash}, averaged over $T_c$ and $T_n$.
According to the above readings of these ungrounded trajectories, the coding agents show their strong statistical prior towards \textsf{no-crash} when observing no intervention site. This prior inflates correct rate by giving ungrounded \textsf{no-crash} verdicts in $T_n$.

\begin{table}[t]
\centering
\small
\setlength{\tabcolsep}{2pt}
\begin{tabular*}{\columnwidth}{@{\extracolsep{\fill}}lcccc@{}}
\toprule
Model
& \makecell[b]{DeepSeek\\V4 Flash}
& \makecell[b]{GLM\\5.3}
& \makecell[b]{Gemini\\3.8 Flash}
& \makecell[b]{Claude\\Opus 5} \\
\midrule
\multicolumn{5}{@{}c}{\textbf{On $T_c$}} \\
\midrule
$\Pr(O_\Delta{=}\mathsf{F} \mid \textsf{no-crash})$ & 49.3 & 38.9 & 31.9 & 34.0 \\
$\Pr(\textsf{no-crash} \mid O_\Delta{=}\mathsf{F})$ & 86.4 & 97.7 & 95.8 & 90.0 \\
\midrule
\multicolumn{5}{@{}c}{\textbf{On $T_n$}} \\
\midrule
$\Pr(O_\Delta{=}\mathsf{F} \mid \textsf{no-crash})$ & 44.0 & 23.8 & 16.9 & 11.3 \\
$\Pr(\textsf{no-crash} \mid O_\Delta{=}\mathsf{F})$ & 87.0 & 97.1 & 95.5 & 91.9 \\
\midrule
\multicolumn{5}{@{}c}{\textbf{Average of $T_c$ and $T_n$}} \\
\midrule
$\Pr(O_\Delta{=}\mathsf{F} \mid \textsf{no-crash})$ & 46.7 & 31.4 & 24.4 & 22.6 \\
$\Pr(\textsf{no-crash} \mid O_\Delta{=}\mathsf{F})$ & 86.7 & 97.4 & 95.7 & 90.9 \\
\bottomrule
\end{tabular*}
\vspace{-0.5\baselineskip}
\caption{Frequency (\%) of coding agents' trajectories in different patterns}
\label{tab:nocrash-decomp}
\vspace{-8pt}
\end{table}

\begin{table}[t]
\centering
\small
\begin{tabular*}{0.92\columnwidth}{@{\extracolsep{\fill}}lrrr@{}}
\toprule
& \multicolumn{2}{c}{$\Pr(v{=}\textsf{crash}\mid\cdot)$ (\%)} & Difference \\
\cmidrule(lr){2-3}
Source tree & $\mathsf{FFTT}$ & $\mathsf{FFFF}$ & (pp.) \\
\midrule
$T_c$ & 42.6 & 6.0 & $\approx37$ \\
$T_n$ & 39.2 & 5.1 & $\approx34$ \\
\midrule
$T_c-T_n$ & $\approx3.5$ & $\approx0.8$ & $\approx3$ \\
\bottomrule
\end{tabular*}
\caption{Opus~5 crash-verdict rates conditioned on $(O_\Delta,D_\Delta,O_C,D_C) \in \{\mathsf{FFTT}, \mathsf{FFFF}\}$}
\label{tab:fftt-ffff-opus}
\vspace{-8pt}
\end{table}

\parabf{Crash-site analysis alone does not ensure discrimination.}
\Cref{tab:fftt-ffff-opus} compares Opus trajectories with profiles $\mathsf{FFTT}$ and $\mathsf{FFFF}$.
On $T_c$, the crash-verdict rate is $42.6\%$ for $\mathsf{FFTT}$, compared with $6.0\%$ for $\mathsf{FFFF}$.
On $T_n$, the corresponding rates are $39.2\%$ and $5.1\%$.
The roughly $37$ pp. increase on $T_c$ is therefore accompanied by a $34$ pp. increase on $T_n$, so discriminability gains only about $3$ pp.
Claude Opus~5 thus predicts a crash more often when it reads a suspicious crash site despite missing the intervention site, and it does so on both $T_c$ and $T_n$.
Results for the other agents are reported in \Cref{tab:fftt-ffff-all} in the Appendix.

\parabf{Source interventions reduce evaluation bias.}
These results show that single-tree accuracy can reward prediction bias rather than crash reasoning.
In paired evaluation, prediction tendencies shared by both trees cancel in $J$ in expectation.
A high $J$ therefore reflects distinguishing execution outcomes, rather than defaulting to \textsf{no-crash} without seeing the intervention site or merely recognizing a potential crash site.




\section{Conclusion}

In this work, we evaluate the crash reasoning of coding agents with our paired evaluation methodology on \sysrb, which pairs source trees with opposite crash outcomes.
The main failure pattern of agents is missing real crashes ($27.5$--$76.8\%$ false-negative rate).
Finding the intervention site is not enough: even when it observes the site, the most capable agent, Claude Opus 5, falls $24.6$~pp.\ short of perfect discriminability.
Discriminability also drops sharply when the crash occurs far from the intervention site.
Finally, single-source-tree evaluations can reward misleading predictions; our paired evaluation reduces this bias, motivating the paired design over single-tree evaluation.

\section*{Limitations}
\label{sec:limit}


\parabf{Necessity check.}
Though the site profile provides a systematic way to highlight the trajectories without grounding on intervention site or crash site, our approach still lacks direct measurement of crash reasoning.

\parabf{Environment mismatch.}
Practitioners debugging system crashes utilize debuggers, git history and online documents to help understand the crash mechanism, while in our experiments we restrict coding agents to local source code navigation and bash commands to limited builtin software.

\parabf{Evaluation cost.}
It requires $K$ runs (usually $\geq 8$) to soundly estimate crash-verdict frequency on a single source tree, and considering the complex nature of system-level crash reasoning, each agentic trajectory is longer than normal tasks.
Therefore, our evaluation can be highly constrained by budget.

\parabf{Source of $\Delta$.}
We only pick the commits from Linux and open-source system library repositories, while mutating code and other methods can generate interventions with opposite execution outcomes.
Given the commits are public on the Internet, coding agents may have been trained on the commits and can produce crash verdict affected by data contamination.

\section*{Ethical Considerations}

We do not foresee significant ethical risks from this work.
All crashes in \textsc{SysRB} come from public sources: syzbot reports, whose reproducers and crash reports are publicly available, and OSS-Fuzz crashes collected by ARVO after they were fixed and disclosed.
The benchmark therefore contains no previously undisclosed vulnerabilities.
The task asks an agent whether a known input crashes a given source tree and, if so, where; it does not involve exploit development or attacks on deployed systems.
All source trees come from open-source projects and remain subject to their original licenses.
The study involves no human subjects or personal data: the manual validation of the trajectory annotator was performed by the authors.


\bibliography{custom}

\appendix

\begin{table*}[t]
\centering
\small
\setlength{\tabcolsep}{4pt}
\begin{tabular}{l l l l l l}
\toprule
Model & Agent & Serving & Weights & Context / sampling \\
\midrule
DeepSeek V4 Flash (0731) & Claude Code 2.1.235 & vLLM, $8\times$B200 & FP8 & 262k; $T{=}1.0$, top-$p$ 0.95 \\
GLM 5.3                  & Claude Code 2.1.235 & vLLM, $8\times$B200 & NVFP4 & 262k; $T{=}1.0$, top-$p$ 0.95 \\
Gemini 3.8 Flash         & Antigravity CLI 1.2.1 & --- & --- & provider default; effort high \\
Claude Opus 5            & Claude Code 2.1.235 & --- & --- & provider default; effort high \\
\bottomrule
\end{tabular}
\caption{Serving and sampling configuration. Both self-hosted models use tensor parallelism 8 with
expert parallelism and lossless speculative decoding; Claude Code compacts its context at about
165k tokens. Effort ``high'' is each CLI's default.}
\label{tab:serving}
\end{table*}

\begin{table*}[t]
\centering

\small
\setlength{\tabcolsep}{3.5pt}

\resizebox{\textwidth}{!}{
\begin{tabular}{l l cc cc cc cc ccc cc}
\toprule
Model & Split & $p_{\mathrm{c}}$ & $p_{\mathrm{n}}$ & $\mathrm{TPR}_{1}$ & $\mathrm{FPR}_{1}$ & $\mathrm{TPR}_{0}$ & $\mathrm{FPR}_{0}$ & $J_{1}$ & $\epsilon$ & $J_\alpha$ & $J_\beta$ & $\kappa$ & $J'$ & $J$ \\
\midrule
DeepSeek V4 Flash (0731)  & Full  & 56.2 & 54.9 & 30.8 & 9.4  & 13.6 & 13.0 & $+21.5$ & $+0.6$ & $+11.9$ & $+0.3$ & $+0.1$ & $+12.3$ & $+12.3$ \\
                          & Linux & 45.9 & 46.1 & 25.7 & 12.1 & 15.9 & 15.8 & $+13.6$ & $+0.2$ & $+6.3$  & $+0.1$ & $-0.0$ & $+6.3$  & $+6.5$  \\
                          & OSS-Fuzz  & 70.4 & 67.1 & 35.4 & 6.8  & 7.8  & 6.9  & $+28.6$ & $+0.9$ & $+19.7$ & $+0.3$ & $+0.5$ & $+20.4$ & $+20.5$ \\
\addlinespace
GLM 5.3                   & Full  & 76.8 & 76.5 & 53.7 & 4.4  & 2.3  & 2.9  & $+49.2$ & $-0.6$ & $+37.7$ & $-0.1$ & $+0.1$ & $+37.7$ & $+37.6$ \\
                          & Linux & 68.5 & 68.2 & 43.0 & 3.3  & 2.0  & 3.2  & $+39.6$ & $-1.1$ & $+27.1$ & $-0.4$ & $+0.0$ & $+26.8$ & $+26.6$ \\
                          & OSS-Fuzz  & 88.4 & 88.0 & 65.3 & 5.6  & 3.2  & 1.9  & $+59.7$ & $+1.2$ & $+52.6$ & $+0.1$ & $+0.1$ & $+52.9$ & $+53.1$ \\
\addlinespace
Gemini 3.8 Flash          & Full  & 83.0 & 82.8 & 58.2 & 2.5  & 4.2  & 4.5  & $+55.7$ & $-0.3$ & $+46.2$ & $-0.0$ & $+0.1$ & $+46.2$ & $+46.2$ \\
                          & Linux & 77.1 & 76.1 & 50.7 & 2.5  & 4.8  & 4.8  & $+48.2$ & $-0.0$ & $+37.0$ & $-0.0$ & $+0.2$ & $+37.2$ & $+37.2$ \\
                          & OSS-Fuzz  & 92.1 & 93.0 & 67.7 & 2.5  & 1.3  & 2.5  & $+65.2$ & $-1.2$ & $+60.4$ & $-0.1$ & $-0.3$ & $+60.0$ & $+60.0$ \\
\addlinespace
Claude Opus 5             & Full  & 89.6 & 88.3 & 79.7 & 4.3  & 10.0 & 8.1  & $+75.4$ & $+1.9$ & $+67.1$ & $+0.2$ & $+0.4$ & $+67.7$ & $+67.7$ \\
                          & Linux & 85.1 & 83.9 & 77.6 & 4.7  & 11.4 & 9.1  & $+72.9$ & $+2.3$ & $+61.6$ & $+0.4$ & $+0.4$ & $+62.3$ & $+62.3$ \\
                          & OSS-Fuzz  & 95.9 & 94.5 & 82.4 & 3.9  & 3.2  & 4.2  & $+78.5$ & $-1.0$ & $+74.8$ & $-0.0$ & $+0.5$ & $+75.2$ & $+75.3$ \\
\bottomrule
\end{tabular}
}
\caption{Full decomposition of $J$ by the $O_\Delta$ gate on Linux + OSS-Fuzz, all values in percent.}
\label{tab:decomp-odelta-full}
\end{table*}

\begin{table*}[t]
\centering
\small
\begin{tabular}{llrrr}
\toprule
Model & Source tree & $P(\text{crash}\mid\mathsf{FFTT})$ & $P(\text{crash}\mid\mathsf{FFFF})$ & Lift \\
 & & (\%) & (\%) & (pp.) \\
\midrule
GLM 5.3 & $T_c$ & 10.5 & 1.6 & +8.8 \\
 & $T_n$ & 13.8 & 2.1 & +11.6 \\
\cmidrule(l){2-5}
 & $T_c - T_n$ & $-$3.3 & $-$0.5 & $-$2.8 \\
\cmidrule(l){1-5}
\addlinespace[3pt]
DeepSeek V4 Flash & $T_c$ & 15.4 & 13.3 & +2.1 \\
 & $T_n$ & 17.8 & 12.6 & +5.2 \\
\cmidrule(l){2-5}
 & $T_c - T_n$ & $-$2.4 & +0.7 & $-$3.1 \\
\cmidrule(l){1-5}
\addlinespace[3pt]
Gemini 3.8 Flash & $T_c$ & 27.0 & 2.5 & +24.6 \\
 & $T_n$ & 26.3 & 2.9 & +23.4 \\
\cmidrule(l){2-5}
 & $T_c - T_n$ & +0.8 & $-$0.4 & +1.2 \\
\cmidrule(l){1-5}
\addlinespace[3pt]
Claude Opus~5 & $T_c$ & 42.6 & 6.0 & +36.7 \\
 & $T_n$ & 39.2 & 5.1 & +34.1 \\
\cmidrule(l){2-5}
 & $T_c - T_n$ & +3.5 & +0.8 & +2.6 \\
\bottomrule
\end{tabular}
\caption{Crash-verdict rates conditioned on $(O_\Delta,D_\Delta,O_C,D_C) \in \{\mathsf{FFTT}, \mathsf{FFFF}\}$}
\label{tab:fftt-ffff-all}
\end{table*}

\section{Experiment Details}

\subsection{Model Serving and Sampling Parameters}
\label{appendix:model-params}

All agents run the same harness, prompt, source tree images and limits
($k=8$ samples per instance and source tree, 2-hour wall clock, no network). The coding CLIs send no
sampling parameters, so each model decodes with its checkpoint's or its provider's defaults.
\Cref{tab:serving} lists the per-model configuration.

When running Claude Opus 5 with Claude Code, we observe rejections to finish the crash reasoning task due to cybersecurity classifier. These trajectories' verdicts are regarded as invalid, distinguished from \textsf{crash} and \textsf{no-crash} verdicts and excluded from our statistics. Same as the ungraded trajectories, we exclude them from our accounting.

\section{Crash Report}
\label{appendix:crash-report}

\paragraph{Crash type and subsystem.}
Both labels are derived deterministically from the dataset with frozen, versioned rule tables, never by a model.
The reference crash type is the first match of an ordered regular-expression table over a closed vocabulary:
for Linux, applied to the syzbot title and then the report head (KASAN, KMSAN, KFENCE, UBSAN, WARNING, BUG,
NULL dereference, general protection fault, deadlock, hung task, memory leak, divide error, stack overflow,
panic); for OSS-Fuzz, applied to ARVO's crash label, its summary, and the sanitizer \texttt{ERROR:} line
(heap/stack/global buffer overflow, heap use-after-free, use-of-uninitialized-value, SEGV, double free,
index out of bounds, \ldots). The agent's \texttt{crashType} is normalized through the same vocabulary and
compared for equality; labels outside it are mismatches, and classes the agent is not offered map to
\emph{other}. The subsystem of a frame is a fixed function of its path: the leading directory
components for Linux (\texttt{mm/huge\_memory.c} $\to$ \texttt{mm}, \texttt{drivers/net/\ldots} $\to$
\texttt{drivers/net}) and the top-level source directory for OSS-Fuzz. The ``same subsystem'' stack score
compares the predicted and reference crash frames' subsystems after stripping compiler suffixes and removing
entry, sanitizer and reporting frames.

\section{$J$ Decomposition}
\label{app:j-decomposition}

\parabf{Setup.}
Fix an agent and a set of instances. Every submitted run carries an source tree label
$Y \in \{\mathrm{c}, \mathrm{n}\}$ (crashing or non-crashing), a verdict
$V \in \{\textsf{crash}, \textsf{no-crash}\}$, and a binary intervention-site gate
$G \in \{0, 1\}$; in the main text $G = \mathbf{1}[O_\Delta{=}\mathsf{T}]$.
The grader leaves $G$ undetermined for a small fraction of runs (an incomplete
transcript, or a judge claim it could not verify); such runs are excluded from
the decomposition population, which is therefore the submitted runs with a
determined gate.
Write $\Pr\nolimits_{\mathrm{c}}[\cdot]$ and $\Pr\nolimits_{\mathrm{n}}[\cdot]$ for the empirical
distributions over the population on each source tree, every run counted once. Then

\begin{equation}
\begin{aligned}
J' &= \mathrm{TPR} - \mathrm{FPR} \\
      &= \Pr\nolimits_{\mathrm{c}}[V{=}\text{crash}] - \Pr\nolimits_{\mathrm{n}}[V{=}\text{crash}].
\end{aligned}
\label{eq:app-j-def}
\end{equation}

Define the per-source tree intervention-site observation rates and the gated and blind verdict rates,
\begin{equation}
\begin{aligned}
p_{\mathrm{c}} &= \Pr\nolimits_{\mathrm{c}}[G{=}1], \quad
p_{\mathrm{n}} = \Pr\nolimits_{\mathrm{n}}[G{=}1], \\
\mathrm{TPR}_{1} &= \Pr\nolimits_{\mathrm{c}}[V{=}\text{crash}\mid G{=}1], \\
\mathrm{TPR}_{0} &= \Pr\nolimits_{\mathrm{c}}[V{=}\text{crash}\mid G{=}0], \\
\mathrm{FPR}_{1} &= \Pr\nolimits_{\mathrm{n}}[V{=}\text{crash}\mid G{=}1], \\
\mathrm{FPR}_{0} &= \Pr\nolimits_{\mathrm{n}}[V{=}\text{crash}\mid G{=}0].
\end{aligned}
\label{eq:app-gate-defs}
\end{equation}

and the conditional discriminabilities
$J_{1} = \mathrm{TPR}_{1} - \mathrm{FPR}_{1}$ (in the main text
$J_{O_\Delta{=}\mathsf{T}}$) and $\epsilon = \mathrm{TPR}_{0} - \mathrm{FPR}_{0}$.

\parabf{Exact identity.}
By the law of total probability on each source tree separately,
$\mathrm{TPR} = p_{\mathrm{c}}\mathrm{TPR}_{1} + (1-p_{\mathrm{c}})\mathrm{TPR}_{0}$
and
$\mathrm{FPR} = p_{\mathrm{n}}\mathrm{FPR}_{1} + (1-p_{\mathrm{n}})\mathrm{FPR}_{0}$,
so
\begin{equation}
\begin{aligned}
J' &= \big[p_{\mathrm{c}}\mathrm{TPR}_{1} - p_{\mathrm{n}}\mathrm{FPR}_{1}\big] \\
   &\quad + \big[(1-p_{\mathrm{c}})\mathrm{TPR}_{0} - (1-p_{\mathrm{n}})\mathrm{FPR}_{0}\big].
\end{aligned}
\label{eq:app-j-exact}
\end{equation}
The two brackets are the contributions of gated and blind runs to $J'$. They are
not multiples of $J_{\alpha}$ and $\epsilon$ in general, because the two source trees are
weighted by their own intervention-site observation rates.

\parabf{Isolating the correction.}
Let $\bar p = \tfrac12(p_{\mathrm{c}} + p_{\mathrm{n}})$ be the source tree-averaged
intervention-site observation rate (the $p$ of the main text) and $\delta = p_{\mathrm{c}} - p_{\mathrm{n}}$,
so $p_{\mathrm{c}} = \bar p + \delta/2$ and $p_{\mathrm{n}} = \bar p - \delta/2$.
Substituting into \eqref{eq:app-j-exact} and collecting terms,
\begin{align*}
&p_{\mathrm{c}}\mathrm{TPR}_{1} - p_{\mathrm{n}}\mathrm{FPR}_{1} \\
&\qquad = \bar p\,J_{1} + \tfrac{\delta}{2}(\mathrm{TPR}_{1}+\mathrm{FPR}_{1}), \\
&(1-p_{\mathrm{c}})\mathrm{TPR}_{0} - (1-p_{\mathrm{n}})\mathrm{FPR}_{0} \\
&\qquad = (1-\bar p)\epsilon - \tfrac{\delta}{2}(\mathrm{TPR}_{0}+\mathrm{FPR}_{0}).
\end{align*}
which gives the exact decomposition
\begin{equation}
\begin{aligned}
J' &= \underbrace{\bar p\,J_{1}}_{J_\alpha}
    +\underbrace{(1-\bar p)\epsilon}_{J_\beta}
    +\kappa,\\
\kappa &= \tfrac{\delta}{2}\big[(\mathrm{TPR}_{1}+\mathrm{FPR}_{1}) \\
       &\qquad - (\mathrm{TPR}_{0}+\mathrm{FPR}_{0})\big].
\end{aligned}
\label{eq:app-j-decomp}
\end{equation}
The first two terms are \Cref{tab:decomp-odelta}; $\kappa$ is exactly what it
omits. $\kappa$ vanishes when the gate fires equally often on both source trees
($\delta = 0$) or when gated and blind runs say ``crash'' at the same overall
rate, and it is bounded by $|\kappa| \le |\delta|$ since each bracket lies in
$[0, 2]$.

\end{document}